\documentclass[lettersize,onecolumn]{IEEEtran}
\usepackage{amsmath,amssymb,amsfonts}
\usepackage{algorithmic}
\usepackage{algorithm}
\usepackage{array}
\usepackage[caption=false,font=normalsize,labelfont=sf,textfont=sf]{subfig}
\usepackage{textcomp}
\usepackage{stfloats}
\usepackage{url}
\usepackage{verbatim}
\usepackage{graphicx}
\usepackage{cite}
\makeatletter
\let\NAT@parse\undefined
\makeatother
\usepackage{hyperref}  
\newtheorem{theorem}{Theorem}
\newtheorem{lemma}{Lemma}
\newtheorem{corollary}{Corollary}
\newtheorem{proposition}[theorem]{Proposition}%

\newtheorem{example}{Example}%
\newtheorem{remark}{Remark}%

\begin{document}
	
	\title{The minimum distance of the antiprimitive BCH code with designed distance 3}
	
	\author{Haojie Xu, Xia Wu,
		Wei Lu and Xiwang Cao
		\thanks{The work of Haojie~Xu was supported by Postgraduate Research $\And$ Practice Innovation Program of Jiangsu Province under Grant KYCX25\_0417. The work of Xia Wu and Wei Lu were supported by the National Natural Science Foundation of China under Grant 12371035, and Jiangsu Provincial Scientific Research Center of Applied Mathematics under Grant BK20233002. The work of Xiwang Cao was supported by the National Natural Science Foundation of China under Grant 12171241. \textit{(Corresponding author: Xia Wu.)}}
		\thanks{Haojie Xu, Xia Wu and Wei Lu are with the School of Mathematics, Southeast University, Nanjing 210096, China. (e-mail: xuhaojiechn@163.com; wuxia80@seu.edu.cn; 
			luwei1010@seu.edu.cn)}
		\thanks{Xiwang Cao is with the Department of Math, Nanjing University of Aeronautics and Astronautics, Nanjing 211100, China. (e-mail: xwcao@nuaa.edu.cn)}}
	
	
	
	\maketitle
	
	\begin{abstract}
		Let $\mathcal{C}_{(q,q^m+1,3,h)}$ denote the antiprimitive BCH code with designed distance 3. In this paper, we demonstrate that the minimum distance $d$ of $\mathcal{C}_{(q,q^m+1,3,h)}$ equals 3 if and only if $\gcd(2h+1,q+1,q^m+1)\ne1$. When both $q$ and $m$ are odd, we determine the sufficient and necessary condition for $d=4$ and fully characterize the minimum distance in this case. Based on these conditions, we investigate the parameters of $\mathcal{C}_{(q,q^m+1,3,h)}$ for certain $h$. Additionally, two infinite families of distance-optimal codes and several linear codes with the best known parameters are presented.
	\end{abstract}
	
	\begin{IEEEkeywords}
		Cyclic code, BCH code, LCD code, antiprimitive BCH code, subfield subcode..
	\end{IEEEkeywords}
	
	\section{Introduction}\label{sec1}
	
	Let $q$ be the power of a prime $p$ and $\mathbb{F}_q$ be the finite field with $q$ elements. An $[n,k]$ {\it linear code} $\mathcal{C}$ over $\mathbb{F}_q$ is a vector subspace of $\mathbb{F}_q^n$ with dimension $k$. The {\it minimum distance} $d$ of a linear code $\mathcal{C}$ is represented by $d=\min\{\mathrm{wt}(\mathbf{c}):\mathbf{c}\in\mathcal{C}\setminus \{\mathbf{0}\}\}$, where $\mathrm{wt}(\mathbf{c})$ is the number of nonzero coordinates of $\mathbf{c}$ and called the {\it Hamming weight} of $\mathbf{c}$. The \textit{dual code} $\mathcal{C}^\perp$ of $\mathcal{C}$ is defined as 
	\begin{equation*}
		\mathcal{C}^\perp=\left\{\mathbf{v}\in\mathbb{F}_q^n:\mathbf{c}\cdot\mathbf{v}=0,\ \forall\;\mathbf{c}\in\mathcal{C}\right\},
	\end{equation*}
	where $\mathbf{c}\cdot\mathbf{v}$ is the inner product of $\mathbf{c}$ and $\mathbf{v}$.
	
	An $[n,k]$ code $\mathcal{C}$ over $\mathbb{F}_q$ is called \textit{cyclic} if $\mathbf{c}=(c_0,c_1,\dots,c_{n-2},c_{n-1})\in\mathcal{C}$ implies $(c_{n-1},c_0,c_1,\dots,c_{n-2})\in\mathcal{C}$. Through the isomorphism $(c_0,c_1,\cdots,c_{n-1})\leftrightarrow c_0+c_1x+\cdots+c_{n-1}x^{n-1}$, the residue class ring $\mathbb{F}_q[x]/(x^n-1)$ is isomorphic to $\mathbb{F}_q^n$ as a vector space over $\mathbb{F}_q$. Then we can regard any linear code $\mathcal{C}$ as a subset of $\mathbb{F}_q[x]/(x^n-1)$. According to the definition of cyclic codes, the linear code $\mathcal{C}$ is cyclic if and only if $\mathcal{C}$ is an ideal of $\mathbb{F}_q[x]/(x^n-1)$. Since every ideal of $\mathbb{F}_q[x]/(x^n-1)$ is principal, every non-zero ideal $\mathcal{C}$ is generated by the unique monic polynomial $g(x)\in\mathbb{F}_q[x]$ of the lowest degree, where $g(x)$ is a factor of $x^n-1$. Suppose that $\mathcal{C}=\left<g(x)\right>$ is a cyclic code. Then $g(x)$ is called the \textit{generator polynomial} of $\mathcal{C}$ and $h(x)=(x^n-1)/g(x)$ is called the \textit{parity-check polynomial} of $\mathcal{C}$.
	
	Throughout this paper, we assume that $\gcd(n,q)=1$. Let $m$ be the multiplicative order of $q$ modulo $n$. Let $\alpha$ be a generator of $\mathbb{F}_{q^m}$ and $\beta=\alpha^{(q^m-1)/n}$. Then $\beta$ is a primitive $n$-th root of unity in $\mathbb{F}_{q^m}$. Denote the \textit{minimal polynomial} over $\mathbb{F}_q$ of $\beta^s$ by $g_s(x)$. Let $h$ be a nonnegative integer and $\delta$ be an integer with $2\le\delta\le n$. Then a \textit{BCH code} over $\mathbb{F}_q$ of length $n$ and \textit{designed distance} $\delta$, denoted by $\mathcal{C}_{(q,n,\delta,h)}$, is a cyclic code generated by 
	\begin{equation*}
		g_{(q,n,\delta,h)}=\mathrm{lcm}\left(g_h(x),g_{h+1}(x),\dots,g_{h+\delta-2}(x)\right),
	\end{equation*}
	where $\mathrm{lcm}(\cdot)$ denotes the least common multiple computed over $\mathbb{F}_q$. If $h=1$, the code $\mathcal{C}_{(q,n,\delta,h)}$ is called a \textit{narrow-sense} BCH code. If $n=q^m-1$, $\mathcal{C}_{(q,n,\delta,h)}$ is called \textit{primitive}. If $n=q^m+1$, then $\mathcal{C}_{(q,n,\delta,h)}$ is referred to as the \textit{antiprimitive} BCH code. According to the BCH bound, the minimum distance of the BCH code $\mathcal{C}_{(q,n,\delta,h)}$ is at least $\delta$.
	
	BCH codes are a crucial class of linear codes due to their exceptional error-correcting capabilities, and simple encoding and decoding algorithms. Over the past several decades, BCH codes have been the subject of extensive investigation, yet their parameters are determined only in a limited number of special cases (see, e.g., \cite{Ding2017DimensionMinimum,Liu2019SomBinary,Liu2022Infinitefamilies,Liu2023DiemnsionNonbinary,Wang2024DualNarrow,Xu2024Thedual,Zhang2024SomeResults}). Notably, all antiprimitive BCH codes are linear complementary dual (LCD for short) codes, which play an important role in resisting side-channel attacks and fault injection attacks \cite{Carlet2015ComplementaryDual}. In recent years, antiprimitive BCH codes have seen rising focus. Li et al. presented a lower bound on the minimum distance of $\mathcal{C}_{(q,q^m+1,\delta,0)}$ \cite{Li2017LCDCyclic}. Ding and Tang obtained near maximum distance separable (NMDS for short) codes through the BCH codes $\mathcal{C}_{(2^m,2^m+1,3,1)}$, $\mathcal{C}_{(3^m,3^m+1,3,1)}$, and $\mathcal{C}_{(2^m,2^m+1,4,1)}$ \cite{Ding2020InfiniteFamiliesMDS,Tang2021InfiniteFamilyLinear}. Geng et al. showed that the BCH code $\mathcal{C}_{(3^m,3^m+1,3,4)}$ with $s$ being odd is an almost MDS (AMDS for short) code \cite{Geng2022ClassAlmostMDS}. Zhu et al. obtained some lower bounds on the minimum distance of antiprimitive BCH codes and several linear codes with good parameters \cite{Zhu2022AryAntiprimitive}. Xu et al. established the sufficient and necessary conditions for the minimum distance of $\mathcal{C}_{(q^m,q^m+1,3,h)}$ to be $3$ and $4$ \cite{Xu2025TheSufficient}.
	
	The main objective of this paper is to study the minimum distance $d$ of the BCH code $\mathcal{C}_{(q,q^m+1,3,h)}$. We establish the sufficient and necessary condition for $d=3$. For odd $q$ and $m$, we further determine the sufficient and necessary condition for $d=4$ and fully determine the minimum distance in this case. Notably, these two findings generalize the results in \cite[Theormes~1 and 3]{Xu2025TheSufficient}. Moreover, we derive the parameters of $\mathcal{C}_{(q,q^m+1,3,h)}$ for certain values of $h$, obtaining two infinite families of distance-optimal linear codes. We also investigate the parameters of $\mathcal{C}_{(2,2^m+1,3,h)}$ and $\mathcal{C}_{(3,3^m+1,3,h)}$, extending the results in \cite[Theorem IV.22]{Zhu2022AryAntiprimitive}. Additionally, several linear codes with the best known parameters are presented.
	
	The rest of this paper is organized as follows. In Section~\ref{sec2}, we provide some concepts of linear codes, $q$-cyclotomic cosets, and elementary number theory. In Section~\ref{sec3}, we establish the sufficient and necessary condition for the minimum distance of $\mathcal{C}_{(q,q+1,3,h)}$ to be $3$ for any $q$ and $m$, and to be $4$ for odd $q$ and $m$. In Section~\ref{sec4}, we present two families of distance-optimal linear codes and several codes with the best known parameters. In Section~\ref{sec5}, we conclude this paper.

	\section{Preliminaries}\label{sec2}
	
	\subsection{Subfield subcode and LCD codes}\label{sec2.1}
	
	Let $\mathcal{C}$ be an $[n,k,d]$ code over $\mathbb{F}_{r^t}$, where $r$ is a prime power. The {\it subfield subcode} $\mathcal{C}|_{\mathbb{F}_r}$ of $\mathcal{C}$ with respect to $\mathbb{F}_r$ is the set of codewords in $\mathcal{C}$ each of whose components is in $\mathbb{F}_r$, i.e., $\mathcal{C}|_{\mathbb{F}_r}=\mathcal{C}\cap\mathbb{F}_r^n$. This is an $[n,k^\prime,d^\prime]$ code with $n-t(n-k)\le k^\prime\le k$ and $d\le d^\prime$ \cite{Delsarte1975SubfieldSubcodes}.
	
	A linear code $\mathcal{C}$ is said to be \textit{linear complementary dual} (LCD for short) if $\mathcal{C}\cap\mathcal{C}^\perp=\{\mathbf{0}\}$. Let $f(x)=f_jx^j+f_{j-1}x^{j-1}+\dots+f_1x+f_0$ be a polynomial over $\mathbb{F}_q$ with $f_j\ne0$ and $f_0\ne0$. The \textit{reciprocal} $f^\ast(x)$ of $f(x)$ is defined by $f^\ast(x)=f_0^{-1}x^jf(x^{-1})$. A polynomial $f(x)$ is called \textit{self-reciprocal} if $f^\ast(x)=f(x)$. A linear code $\mathcal{C}$ is said to be \textit{reversible} if $\mathbf{c}=(c_0,c_1,\dots,c_{n-2},c_{n-1})\in\mathcal{C}$ implies $(c_{n-1},c_{n-2},\dots,c_1,c_0)\in\mathcal{C}$. According to Theorem~\ref{sec2 thm1}, one can deduce that every antiprimitive BCH code over $\mathbb{F}_q$ is an LCD code.
	
	\begin{theorem}\cite{Massey1964ReversibleCodes}, \cite[p.206]{MacWilliams1977TheTheory}, \cite{Yang1994TheCondition}\label{sec2 thm1}
		Let $\mathcal{C}$ be a cyclic code of length $n$ over $\mathbb{F}_q$ with generator polynomial $g(x)$. Then	the following statements are equivalent.
		\begin{itemize}
			\item $\mathcal{C}$ is an LCD code.
			\item $\mathcal{C}$ is reversible.
			\item $g(x)$ is self-reciprocal.
			\item $\beta^{-1}$ is a root of $g(x)$ for every root $\beta$ of $g(x)$ over the
			splitting field of $g(x)$.
		\end{itemize}
		Furthermore, if $-1$ is a power of $q$ mod $n$, then every cyclic code over $\mathbb{F}_q$ of length $n$ is reversible.
	\end{theorem}
	
	\subsection{Bounds for linear codes}
	
	It is desirable to design linear codes with the largest possible rate $\frac{k}{n}$ and minimum distance $d$ in coding theory. Nevertheless, there are some tradeoffs among $n$, $k$, and $d$. The \textit{Singleton bound} indicates that $d\le n-k+1$. Linear codes that meet the Singleton bound are called \textit{maximum distance separable}, or MDS for short. Note that if $\mathcal{C}$ is MDS so is the dual code $\mathcal{C}^\perp$. An $[n,k,n-k]$ code is said to be \textit{almost MDS} (AMDS for short) \cite{Boer1996AlmostMDSCodes}. Unlike MDS code, the dual code of an AMDS code need not be AMDS. Furthermore, $\mathcal{C}$ is said to be \textit{near MDS} (NMDS for short) if both $\mathcal{C}$ and its dual code $\mathcal{C}^\perp$ are AMDS codes \cite{Dodunekov1995NearMDSCodes}. 
	
	Apart from the Singleton bound, the Sphere Packing bound provides another fundamental tradeoff among the parameters of a linear code. 
	
	\begin{theorem} \rm{(Sphere Packing Bound)} \cite[Theorem 1.12.1]{Huffman2003FUndamentals}
		Let $\mathcal{C}$ be an $[n,k,d]$ linear code. Then 
		\begin{equation}\label{sec2 equ1}
			\sum\limits_{i=0}^{\lfloor\frac{d-1}{2}\rfloor}\binom{n}{i}(q-1)^i\le q^{n-k}.
		\end{equation}
	\end{theorem}
	An $[n,k,d]$ linear code over $\mathbb{F}_q$ is referred to as \textit{distance-optimal} (respectively, \textit{dimension-optimal} and \textit{length-optimal}) if there is no $[n,k,d^\prime\ge d+1]$ (respectively, $[n, k^\prime \ge k + 1, d]$ and $[n^\prime\le n-1,k,d]$) linear code over $\mathbb{F}_q$.
	
	\subsection{$q$-cyclotomic coset}\label{sec2.2}
	
	In this subsection, we will introduce $q$-cyclotomic cosets modulo $n$ to handle the generator polynomial $g(x)$ of the BCH code. Let $\mathbb{Z}_n$ denote the set $\{0,1,2,\dots,n-1\}$ and let $s<n$ be a nonnegative integer. The \textit{$q$-cyclotomic coset of $s$ modulo $n$} is defined as
	\begin{equation}\label{sec2.1 equ1}
		C_s=\{s,sq,sq^2,\ldots,sq^{\ell_s-1}\}\bmod n\subseteq\mathbb{Z}_n,
	\end{equation}
	where $\ell_s$, called the \textit{size} of the $q$-cyclotomic coset, is the smallest positive integer such that $s \equiv sq^{\ell_s} \bmod n$. Then the minimal polynomial over $\mathbb{F}_q$ of $\beta^s$ is
	\begin{equation}\label{sec2.1 equ3}
		g_s(x)=\prod\limits_{i\in C_s}(x-\beta^i)\in\mathbb{F}_q[x].
	\end{equation}
	The smallest integer in $C_s$ is called the \textit{coset leader} of $C_s$. Let $\Gamma_{(n,q)}$ be the set of all the coset leaders. Then $C_s \cap C_t = \emptyset$ for any two distinct elements $s$ and $t$ in $\Gamma_{(n,q)}$. The following result will be useful for calculating $\ell_s$.
	
	\begin{theorem}\cite[Theorem 4.1.4]{Huffman2003FUndamentals}\label{sec2 thm2}
		The size $\ell_s$ of each $q$-cyclotomic coset $C_s$ is a divisor of the size $\ell_1$ of $C_1$.
	\end{theorem}
	
	\subsection{Two results in number theory}\label{sec2.3}
	
	The following are two results in elementary number theory, which will be useful for calculating the dimensions of BCH codes.
	
	\begin{lemma}\cite[Lemma~2]{Xu2024Thedual}\label{sec2.3 lem1}
		Let $p,i$ and $s$ be three positive integers. Suppose that $\gcd(i,s)=m$. Then
		\begin{equation*}
			\gcd(p^i+1,p^s+1)=\left\{\begin{array}{ll}
				p^m+1 &\mathrm{if}\ \frac{i}{m}\ \mathrm{and}\ \frac{s}{m}\ \mathrm{are\ odd},\\
				\begin{array}{ll}
					1&\mathrm{if}\ p\ \mathrm{is\ even,}\\
					2&\mathrm{if}\ p\ \mathrm{is\ odd,}\\
				\end{array} &\mathrm{otherwise}.\\
			\end{array}\right.
		\end{equation*}
	\end{lemma}
	
	\begin{lemma}\cite[Lemma 2.6]{Coulter1998Explicit}\cite[Lemma 2.1]{Coulter1999Evaluation}
		\label{sec2.3 lem2}
		Let $p,i$ and $s$ be three positive integers. Suppose that $\gcd(i,s)=m$. Then
		\begin{equation*}
			\gcd(p^i-1,p^s+1)=\left\{\begin{array}{ll}
				p^m+1 &\mathrm{if}\ \frac{i}{m}\ \mathrm{is\ even},\\
				\begin{array}{ll}
					1&\mathrm{if}\ p\ \mathrm{is\ even,}\\
					2&\mathrm{if}\ p\ \mathrm{is\ odd,}\\
				\end{array} &\mathrm{otherwise}.\\
			\end{array}\right.
		\end{equation*}
	\end{lemma}

	\section{The conditions for the minimum distance of $\mathcal{C}_{(q,q^m+1,3,h)}$ to be $3$ and $4$}\label{sec3}
	Throughout this section, let $q$ be the power of a prime $p$ and $m$ be a positive integer. Let $0\le h\le q^m$ be a positive integer, and $U_l$ be the set of all $l$-th roots of unity in $\mathbb{F}_{q^{2m}}$. It should be noted that for any $x\in U_{q^m+1}$ a straightforward yet crucial property is $x^{q^m}=x^{-1}$. Denote the minimum distance of the BCH code $\mathcal{C}_{(q,q^m+1,3,h)}$ by $d$. In this section, we provide the conditions for $d$ to be 3 and 4, respectively.
	
	Before proving Theorem~\ref{sec3 thm1}, we display a necessary lemma whose proof is straightforward. 
	\begin{lemma}\label{sec3 lem1}
		Define
		\begin{equation}\label{sec3 lem1 equ1}
			D(x,y)=\left|\begin{array}{ll}
				x^h & y^h\\
				x^{h+1} & y^{h+1}
			\end{array}\right|=x^hy^h(y-x),
		\end{equation}
		where $x,y\in\mathbb{F}_{q^{2m}}$. If $x,y\in U_{q^m+1}$, then $D(x,y)^{q^m}=-x^{-2h-1}y^{-2h-1}D(x,y)$.
	\end{lemma}
	
	The following theorem presents the sufficient and necessary condition for $d=3$.
	\begin{theorem}\label{sec3 thm1}
		The minimum distance $d$ of $\mathcal{C}_{(q,{q^m}+1,3,h)}$ is equal to 3 if and only if $\gcd(2h+1,q+1,{q^m}+1)\ne1$. Equivalently, $d\ge4$ if and only if $\gcd(2h+1,q+1,{q^m}+1)=1$.
	\end{theorem}
	\begin{IEEEproof}
		By the BCH bound, we have $d\ge3$. Let $\alpha$ be a generator of $\mathbb{F}_{q^{2m}}^{\ast}$ and $\beta=\alpha^{{q^m}-1}$. Then $\beta$ is a primitive $({q^m}+1)$-th root of unity in $\mathbb{F}_{q^{2m}}$. Define
		\begin{equation*}
			\left.H=\left[\begin{array}{lllll}1&\beta^{h}&(\beta^{h})^2&\cdots&(\beta^{h})^{q^m}\\1&\beta^{h+1}&(\beta^{h+1})^2&\cdots&(\beta^{h+1})^{q^m}\end{array}\right.\right].
		\end{equation*}
		It is easily seen that $H$ is a parity-check matrix of $\mathcal{C}_{(q,{q^m}+1,3,h)}$. Note that $d=3$ if and only if there are three pairwise distinct elements $x, y, z \in U_{{q^m}+1}$ such that
		\begin{equation}
			\label{sec3 thm1 equ1}
			\left.i\left[\begin{array}{l}x^{h}\\x^{h+1}\end{array}\right.\right]+j\left[\begin{array}{l}y^{h}\\y^{h+1}\end{array}\right]-k\left[\begin{array}{l}z^{h}\\z^{h+1}\end{array}\right]=\mathbf{0},
		\end{equation}
		where $i,j,k \in \mathbb{F}_{q}^\ast$. The above equation is the same as
		\begin{equation}
			\label{sec3 thm1 equ2}
			\left.\left[\begin{array}{ll}x^{h} & y^{h}\\x^{h+1} & y^{h+1}\end{array}\right.\right]\left[\begin{array}{l}i\\j\end{array}\right]=\left[\begin{array}{l}kz^{h}\\kz^{h+1}\end{array}\right].
		\end{equation}
		
		First, assume that $d=3$. Since $x\ne y$, it follows from (\ref{sec3 lem1 equ1}) that $D(x,y)\ne0$. Applying Cramer's Rule to (\ref{sec3 thm1 equ2}), we have
		\begin{equation*}
			i=k\frac{D(z,y)}{D(x,y)}\ \mathrm{and}\ j=k\frac{D(x,z)}{D(x,y)}.
		\end{equation*}
		Since $k\in \mathbb{F}_q^\ast$, $k^{q^m}=k$. According to Lemma~\ref{sec3 lem1}, we have
		\begin{equation*}
			i^{q^m}=k\frac{D(z,y)}{D(x,y)}\left(\frac{x}{z}\right)^{2h+1}\ \mathrm{and}\ j^{q^m}=k\frac{D(x,z)}{D(x,y)}\left(\frac{y}{z}\right)^{2h+1}.
		\end{equation*}
		Since $i,j\in\mathbb{F}_q^\ast$, $i^{q^m}=i$, which implies
		\begin{equation*}
			k\frac{D(z,y)}{D(x,y)}\left(\left(\frac{x}{z}\right)^{2h+1}-1\right)=0,
		\end{equation*}
		and so
		\begin{equation}\label{sec3 thm1 equ3}
			\left(\frac{x}{z}\right)^{2h+1}=1.
		\end{equation}
		Similarly, by $j^{q^m}=j$, we obtain
		\begin{equation}\label{sec3 thm1 equ4}
			\left(\frac{y}{z}\right)^{2h+1}=1.
		\end{equation}
		Let $l=\gcd(2h+1,{q^m}+1)$. Since $x,y,z$ are pairwise distinct, $1,\frac{x}{z},\frac{y}{z}$ are also pairwise distinct. Note that $1,\frac{x}{z},\frac{y}{z}\in U_l$. Thus, there are at least three different elements in $U_l$, i.e. $l\ne 1$. Next we will show that $\gcd(l,q+1)\ne1$. Raising to the $q$-th power both sides of (\ref{sec3 thm1 equ1}), we have
		\begin{equation}
			\label{sec3 thm1 equ5}
			\left.i\left[\begin{array}{l}x^{qh}\\x^{q(h+1)}\end{array}\right.\right]+j\left[\begin{array}{l}y^{qh}\\y^{q(h+1)}\end{array}\right]-k\left[\begin{array}{l}z^{qh}\\z^{q(h+1)}\end{array}\right]=\mathbf{0}.
		\end{equation}
		Combining (\ref{sec3 thm1 equ1}) and (\ref{sec3 thm1 equ5}), we get
		\begin{equation*}
			\left|\begin{array}{lll}
				x^{q(h+1)} & y^{q(h+1)} & z^{q(h+1)}\\ x^h & y^h & z^h\\ x^{h+1} & y^{h+1} & z^{h+1}
			\end{array}\right|
			=x^hy^hz^h\left[(x-z)\left(y^{q(h+1)-h}-z^{q(h+1)-h}\right)-(y-z)\left(x^{q(h+1)-h}-z^{q(h+1)-h}\right)\right]=0, 
		\end{equation*}
		which leads to
		\begin{equation}
			\label{sec3 thm1 equ6}
			(x-z)\left(y^{q(h+1)-h}-z^{q(h+1)-h}\right)=(y-z)\left(x^{q(h+1)-h}-z^{q(h+1)-h}\right).
		\end{equation}
		Raising to the $q^m$-th power both sides of (\ref{sec3 thm1 equ6}),
		\begin{equation*}
			(x^{-1}-z^{-1})\left(y^{-q(h+1)+h}-z^{-q(h+1)+h}\right)=(y^{-1}-z^{-1})\left(x^{-q(h+1)+h}-z^{-q(h+1)+h}\right),
		\end{equation*}
		which is the same as
		\begin{equation}\label{sec3 thm1 equ7}
			x^{(q-1)(h+1)}(x-z)\left(y^{q(h+1)-h}-z^{q(h+1)-h}\right)= y^{(q-1)(h+1)}(y-z)\left(x^{q(h+1)-h}-z^{q(h+1)-h}\right)
		\end{equation}
		Now we proof that both sides of (\ref{sec3 thm1 equ6}) are equal to $0$. If neither side of (\ref{sec3 thm1 equ6}) is equal to 0, then we have $x^{(q-1)(h+1)}=y^{(q-1)(h+1)}$ by combining (\ref{sec3 thm1 equ6}) and (\ref{sec3 thm1 equ7}). Then we get \begin{equation*}
			\left(\frac{y}{x}\right)^{(h+1)(q-1)}=1,
		\end{equation*}
		i.e., $\left(\frac{y}{x}\right)^{h+1}\in U_{q-1}$. It follows from (\ref{sec3 thm1 equ3}) and (\ref{sec3 thm1 equ4}) that $(\frac{y}{x})^{2h+1}=1$, so $(\frac{y}{x})^{\gcd(2h+1,(h+1)(q-1),q^m+1)}=1$. Note that $\gcd(q-1,q^m+1)=1$ or $2$. Since $\gcd(2h+1,h+1)=\gcd(2h+1,2(h+1))=1$, we get $\gcd(2h+1,(h+1)(q-1),q^m+1)=1$. Then we have $x=y$, a contradiction. Hence, both sides of (\ref{sec3 thm1 equ6}) equal 0. Since $x\ne z$ and $y\ne z$, we get $x^{q(h+1)-h}=y^{q(h+1)-h}=z^{q(h+1)-h}$, which implies that
		\begin{equation*}
			\left(\frac{x}{z}\right)^{q(h+1)-h}=\left(\frac{x}{z}\right)^{(q+1)(h+1)}=\left(\frac{y}{z}\right)^{q(h+1)-h}=\left(\frac{y}{z}\right)^{(q+1)(h+1)}=1,
		\end{equation*}
		where the first identity holds as (\ref{sec3 thm1 equ3}) and the third identity holds as (\ref{sec3 thm1 equ4}). Thus, $\left(\frac{x}{z}\right)^{h+1},\left(\frac{y}{z}\right)^{h+1}\in U_{q+1}$. Since $\frac{x}{z}, \frac{y}{z}\in U_l$, $\left(\frac{x}{z}\right)^{h+1},\left(\frac{y}{z}\right)^{h+1}\in U_{l}$. Therefore, we have $\gcd(l,q+1)\ne1$, i.e., $\gcd(2h+1,q+1,q^m+1)\ne1$.
		
		Conversely, assume that $r=\gcd(2h+1,q+1,q^m+1)\ne1$, then $r\ge3$ as $2h + 1$ is odd. Suppose that $x,y,z$ are three pairwise distinct elements of $U_r$, then we have $\left(\frac{x}{z}\right)^{2h+1}=\left(\frac{y}{z}\right)^{2h+1}=1$. Take $i:=\frac{D(z,y)}{D(x,y)}, j:=\frac{D(x,z)}{D(x,y)}$ and $k := 1$. Note that $x,y,z\in U_{q+1}$. It then follows from Lemma~\ref{sec3 lem1} that $i^q = i$ and $j^q = j$. According to Cramer’s Rule, (\ref{sec3 thm1 equ2}) is valid. Equivalently, (\ref{sec3 thm1 equ1}) is satisfied, and so $d = 3$.
		
		According to the BCH bound, the contrapositive of the first part is that $d\ge4$ if and only if $\gcd(2h+1,q+1,{q^m}+1)=1$. This completes the proof.
	\end{IEEEproof}
	
	\begin{remark}
		When $m=1$, it was proved in \cite[Theorem~1]{Xu2025TheSufficient} that $d=3$ if and only if $\gcd(2h+1,q+1)\ne1$. It is easy to see that Theorem~\ref{sec3 thm1} generalizes this result.
	\end{remark}
	
	Note that $\gcd(q+1,q^m+1)$ equals $q+1$ if $m$ is odd, and 1 or 2 if $m$ is even. Since $2h+1$ is odd, $\gcd(2h+1,q+1,q^m+1)$ is equal to $\gcd(2h+1,q+1)$ if $m$ is odd, and 1 if $m$ is even. Therefore, Theorem~\ref{sec3 thm1} can be expressed as follows.
	
	\begin{corollary}\label{sec3 cor1}
		The minimum distance $d$ of $\mathcal{C}_{(q,{q^m}+1,3,h)}$ is equal to 3 if and only if $m$ is odd and $\gcd(2h+1,q+1)\ne1$. Equivalently, $d\ge4$ if and only if $m$ is even, or $m$ is odd and $\gcd(2h+1,q+1)=1$.
	\end{corollary}

	Before commencing the proof of Proposition~\ref{sec3 thm2}, we need the following lemma. 
	
	\begin{lemma}\label{sec3 lem2}
		Define
		\begin{equation}\label{sec3 lem2 equ1}
			E(x,y)=\frac{x^{2h+1}-y^{2h+1}}{x-y},
		\end{equation}
		where $x,y\in\mathbb{F}_{q^{2m}}$. If $x,y\in U_{{q^m}+1}$, then $E(x,y)^{q^m}=x^{-2h}y^{-2h}E(x,y)$.
	\end{lemma}
	
	When $\gcd(2h+1,q^m+1)=1$, the following proposition provides a necessary condition for $d=4$.
	\begin{proposition}\label{sec3 thm2}
		Suppose that $\gcd(2h+1,q^m+1)=1$. If the minimum distance $d$ of $\mathcal{C}_{(q,{q^m}+1,3,h)}$ is equal to 4, then there exist four pairwise distinct elements $x,y,z,w\in U_{q^m+1}$ such that
		\begin{equation*}
			\frac{E(x,z)}{E(x,w)}=\frac{E(y,z)}{E(y,w)}.
		\end{equation*}
		Equivalently, if there do not exist four pairwise distinct elements $x,y,z,w\in U_{q^m+1}$ such that $\frac{E(x,z)}{E(x,w)}=\frac{E(y,z)}{E(y,w)}$, then $d\ge5$.
	\end{proposition}
	\begin{IEEEproof}
		Recall from the proof of Theorem~\ref{sec3 thm1} that $\alpha$ is a generator of $\mathbb{F}_{q^{2m}}^{\ast}$, $\beta=\alpha^{{q^m}-1}$ is a primitive $({q^m}+1)$-th root of unity in $\mathbb{F}_{q^{2m}}$, and
		\begin{equation*}
			\left.H=\left[\begin{array}{lllll}1&\beta^{h}&(\beta^{h})^2&\cdots&(\beta^{h})^{q^m}\\1&\beta^{h+1}&(\beta^{h+1})^2&\cdots&(\beta^{h+1})^{q^m}\end{array}\right.\right]
		\end{equation*}
		is a parity-check matrix of $\mathcal{C}_{(q,{q^m}+1,3,h)}$. 
		
		Since $\gcd(2h+1,q^m+1)=1$, we have $d\ge4$ by Theorem~\ref{sec3 thm1}. Note that $d=4$ if and only if there exist four pairwise distinct elements $x,y,z,w\in U_{{q^m}+1}$ such that
		\begin{equation}
			\label{sec3 thm2 equ1}
			\left.i\left[\begin{array}{l}x^{h}\\x^{h+1}\end{array}\right.\right]+j\left[\begin{array}{l}y^{h}\\y^{h+1}\end{array}\right]-k\left[\begin{array}{l}z^{h}\\z^{h+1}\end{array}\right]-l\left[\begin{array}{l}w^{h}\\w^{h+1}\end{array}\right]=\mathbf{0},
		\end{equation}
		where $i,j,k,l\in\mathbb{F}_q^\ast$. Furthermore, (\ref{sec3 thm2 equ1}) is equivalent to
		\begin{equation}
			\label{sec3 thm2 equ2}
			\left.\left[\begin{array}{ll}x^{h} & y^{h}\\x^{h+1} & y^{h+1}\end{array}\right.\right]\left[\begin{array}{l}i\\j\end{array}\right]=\left[\begin{array}{l}kz^{h}+lw^{h}\\kz^{h+1}+lw^{h+1}\end{array}\right].
		\end{equation}
		
		Assume that $d=4$. Since $x\ne y$, it follows from (\ref{sec3 lem1 equ1}) that $D(x,y)\ne0$. Then applying Cramer's Rule to (\ref{sec3 thm2 equ2}) yields
		\begin{equation*}
			i=\frac{\left|\begin{array}{ll}kz^{h}+lw^{h} & y^{h}\\kz^{h+1}+lw^{h+1}&y^{h+1}\end{array}\right|}{D(x,y)}=k\frac{D(z,y)}{D(x,y)}+l\frac{D(w,y)}{D(x,y)}
		\end{equation*}
		and
		\begin{equation*}
			j=\frac{\left|\begin{array}{ll}x^{h} &  kz^{h}+lw^{h}\\x^{h+1} & kz^{h+1}+lw^{n+1}\end{array}\right|}{D(x,y)}=k\frac{D(x,z)}{D(x,y)}+l\frac{D(x,w)}{D(x,y)}.
		\end{equation*}
		Note that $k,l\in \mathbb{F}_q^\ast$, then $k^{q^m}=k$ and $l^{q^m}=l$. According to Lemma~\ref{sec3 lem1}, we have
		\begin{equation*}
			i^{q^m}=k\frac{D(z,y)}{D(x,y)}\left(\frac{x}{z}\right)^{2h+1}+l\frac{D(w,y)}{D(x,y)}\left(\frac{x}{w}\right)^{2h+1}
		\end{equation*}
		and 
		\begin{equation*}
			j^{q^m}=k\frac{D(x,z)}{D(x,y)}\left(\frac{y}{z}\right)^{2h+1}+l\frac{D(x,w)}{D(x,y)}\left(\frac{y}{w}\right)^{2h+1}.
		\end{equation*}
		Since $i,j\in\mathbb{F}_q^\ast$, $i^{q^m}=i$, which implies
		\begin{equation*}
			k\frac{D(z,y)}{D(x,y)}\left(\frac{x}{z}\right)^{2h+1}+l\frac{D(w,y)}{D(x,y)}\left(\frac{x}{w}\right)^{2h+1}=k\frac{D(z,y)}{D(x,y)}+l\frac{D(w,y)}{D(x,y)}.
		\end{equation*}
		Rearranging terms, we get
		\begin{equation}\label{sec3 thm2 equ3}
			k\frac{D(z,y)}{D(x,y)}\left(\left(\frac{x}{z}\right)^{2h+1}-1\right)=-l\frac{D(w,y)}{D(x,y)}\left(\left(\frac{x}{w}\right)^{2h+1}-1\right).
		\end{equation}
		Since $x,y,z,w$ are pairwise distinct and $\gcd(2h+1,{q^m}+1)=1$, $\left(\frac{x}{z}\right)^{2h+1}-1\ne0$ and $\left(\frac{x}{w}\right)^{2h+1}-1\ne0$. Note that $D(z,y)\ne0$ and $D(w,y)\ne0$. Hence, neither side of (\ref{sec3 thm2 equ3}) is equal to 0. Then
		\begin{equation}\label{sec3 thm2 equ4}
			-\frac{k}{l}=\frac{D(w,y)\left(\left(\frac{x}{w}\right)^{2h+1}-1\right)}{D(z,y)\left(\left(\frac{x}{z}\right)^{2h+1}-1\right)}.
		\end{equation}
		Similarly, by $j^{q^m}=j$, we obtain
		\begin{equation}\label{sec3 thm2 equ5}
			-\frac{k}{l}=\frac{D(x,w)\left(\left(\frac{y}{w}\right)^{2h+1}-1\right)}{D(x,z)\left(\left(\frac{y}{z}\right)^{2h+1}-1\right)}.
		\end{equation}
		By combining (\ref{sec3 thm2 equ4}) and (\ref{sec3 thm2 equ5}), we have
		\begin{equation}\label{sec3 thm2 equ6}
			\frac{D(w,y)\left(\left(\frac{x}{w}\right)^{2h+1}-1\right)}{D(z,y)\left(\left(\frac{x}{z}\right)^{2h+1}-1\right)}=\frac{D(x,w)\left(\left(\frac{y}{w}\right)^{2h+1}-1\right)}{D(x,z)\left(\left(\frac{y}{z}\right)^{2h+1}-1\right)}.
		\end{equation}
		Multiplying both sides of (\ref{sec3 thm2 equ6}) by $\frac{w^{2h+1}}{z^{2h+1}}$ yields
		\begin{equation*}
			\frac{E(x,z)}{E(x,w)}=\frac{E(y,z)}{E(y,w)}.
		\end{equation*}
		The proof is completed. 
	\end{IEEEproof}
	
	\begin{remark}
		According to the proof of Theorem~\ref{sec3 thm4} below, when $q$ is odd, there always exist four pairwise distinct elements in $U_{q^m+1}$ satisfying the necessary condition stated in Proposition~\ref{sec3 thm2}. However, the BCH code $\mathcal{C}_{(3,3^2+1,3,1)}$ has minimum distance $5$. Hence, this necessary condition for $d=4$ fails to be sufficient.
	\end{remark}
	
	When $m$ is odd, the following result gives a sufficient condition for $d=4$.
	
	\begin{proposition}\label{sec3 thm3}
		Suppose that $m$ is odd. If $\gcd(2h+1, q+1)=1$, and there exist four pairwise distinct elements $x,y,z,w\in U_{q+1}$ such that $\frac{E(x,z)}{E(x,w)}=\frac{E(y,z)}{E(y,w)}$, then the minimum distance $d$ of $\mathcal{C}_{(q,q^m+1,3,h)}$ is equal to 4.  
	\end{proposition}
	\begin{IEEEproof}
		Assume that $\gcd(2h+1,{q}+1)=1$, then we have $d\ge4$ by Corollary~\ref{sec3 cor1}. Since $m$ is odd, $U_{q+1}\subseteq U_{q^m+1}$. Suppose that there exist four pairwise distinct elements $x,y,z,w\in U_{q+1}$ such that 
		\begin{equation*}
			\frac{E(x,z)}{E(x,w)}=\frac{E(y,z)}{E(y,w)}.
		\end{equation*}
		According to the proof of Proposition~\ref{sec3 thm2}, the above equation is equivalent to (\ref{sec3 thm2 equ6}). Since $\gcd(2h+1,q+1)=1$ and $x,y,z,w\in U_{q+1}$ are different from each other, we can take $l:=-1$ and 
		\begin{equation*}
			k:=\frac{D(w,y)\left(\left(\frac{x}{w}\right)^{2h+1}-1\right)}{D(z,y)\left(\left(\frac{x}{z}\right)^{2h+1}-1\right)}=\frac{D(x,w)\left(\left(\frac{y}{w}\right)^{2h+1}-1\right)}{D(x,z)\left(\left(\frac{y}{z}\right)^{2h+1}-1\right)}\ne0.
		\end{equation*}
		According to Lemma~\ref{sec3 lem1}, we have 
		\begin{equation*}
			k^q=\frac{D(w,y)z^{2h+1}\left(\left(\frac{x}{w}\right)^{q(2h+1)}-1\right)}{D(z,y)w^{2h+1}\left(\left(\frac{x}{z}\right)^{q(2h+1)}-1\right)}=\frac{D(w,y)z^{2h+1}\left(\left(\frac{x}{w}\right)^{-(2h+1)}-1\right)}{D(z,y)w^{2h+1}\left(\left(\frac{x}{z}\right)^{-(2h+1)}-1\right)}=k,
		\end{equation*}
		where the second identity holds since $\frac{x}{w},\frac{x}{z}\in U_{q+1}$. Thus, $k\in\mathbb{F}_q^\ast$. Let
		\begin{equation*}
			i:=k\frac{D(z,y)}{D(x,y)}+l\frac{D(w,y)}{D(x,y)}\ \mathrm{and}\ j:=k\frac{D(x,z)}{D(x,y)}+l\frac{D(x,w)}{D(x,y)}.
		\end{equation*}
		Substituting $k$ and $l$ into $i^q-i$ and $j^q-j$ leads to 
		\begin{equation*}
			i^q-i=k\frac{D(z,y)}{D(x,y)}\left(\left(\frac{x}{z}\right)^{2h+1}-1\right)+l\frac{D(w,y)}{D(x,y)}\left(\left(\frac{x}{w}\right)^{2h+1}-1\right)=0
		\end{equation*}
		and
		\begin{equation*}
			j^q-j=k\frac{D(x,z)}{D(x,y)}\left(\left(\frac{y}{z}\right)^{2h+1}-1\right)+l\frac{D(x,w)}{D(x,y)}\left(\left(\frac{y}{w}\right)^{2h+1}-1\right)=0.
		\end{equation*}
		Thus, we have $i,j\in\mathbb{F}_q$. By the definition of $i,j$, and Cramer's Rule, we have (\ref{sec3 thm2 equ2}) holds. Equivalently, (\ref{sec3 thm2 equ1}) is satisfied. Hence, we have $d=4$. This completes the proof.
	\end{IEEEproof}
	
	\begin{remark}
		When $m$ is even, $\gcd(q+1,q^m+1)$ is equal to $1$ or $2$. Therefore, the conditions described in Proposition~\ref{sec3 thm3} are not valid for $m$ even.
	\end{remark}
	
	When $q$ is even, it is challenging to find four pairwise distinct elements $x,y,z,w\in U_{q+1}$ such that $\frac{E(x,z)}{E(x,w)}=\frac{E(y,z)}{E(y,w)}$. In contrast, when $q$ is odd, the appropriate $x,y,z,w\in U_{q+1}$ in Proposition~\ref{sec3 thm3} always exist.
	\begin{theorem}\label{sec3 thm4}
		Suppose that $q$ and $m$ are odd. Let $d$ be the minimum distance of $\mathcal{C}_{(q,q^m+1,3,h)}$. Then the following hold:
		
		(1) $d=3$ if and only if $\gcd(2h+1,q+1)\ne1$;
		
		(2) $d=4$ if and only if $\gcd(2h+1, q+1)=1$. 
	\end{theorem}
	\begin{IEEEproof}
		Note that $\gcd(q+1,q^m+1)=q+1$ when $m$ is odd. According to Theorem~\ref{sec3 thm1} and Proposition~\ref{sec3 thm3}, we only need to prove that there exist four pairwise distinct elements $x,y,z,w\in U_{q+1}$ such that $\frac{E(x,z)}{E(x,w)}=\frac{E(y,z)}{E(y,w)}$. Since $q$ is odd, $1,-1\in U_{q+1}$ are different elements. Note that $q+1\ge4$. Suppose that $x\in U_{q+1}\setminus\{1,-1\}$. Then we have $x^{-1}\in U_{q+1}$ and $x,x^{-1},1,-1$ are pairwise distinct. Substituting $(x,x^{-1},1,-1)$ for $(x,y,z,w)$ into $\frac{E(y,z)}{E(y,w)}$ yields
		\begin{equation*}
			\frac{E(y,z)}{E(y,w)}=\frac{E(x^{-1},1)}{E(x^{-1},-1)}=\frac{\frac{x^{-2h-1}-1}{x^{-1}-1}}{\frac{x^{-2h-1}+1}{x^{-1}+1}}=\frac{\frac{x^{2h+1}-1}{x-1}}{\frac{x^{2h+1}+1}{x+1}}=\frac{E(x,1)}{E(x,-1)}=\frac{E(x,z)}{E(x,w)}.
		\end{equation*}
		This proof is completed.
	\end{IEEEproof}
	
	\begin{remark}
		By Theorem~\ref{sec3 thm4}, when both $q$ and $m$ are odd, the minimum distance of $\mathcal{C}_{(q,q^m+1,3,h)}$ can be fully characterized. When $m=1$ and $q$ is odd, it was demonstrated in \cite[Theorem~3]{Xu2025TheSufficient} that $d=4$ if and only if $\gcd(2h+1,q+1)=1$. Obviously, this result is covered by Theorem~\ref{sec3 thm4}.
	\end{remark}
	
	When $q$ is odd and $m$ is even, the following proposition presents two sufficient conditions for the minimum distance of $\mathcal{C}_{(q,q^m+1,3,h)}$ to be 4.
	
	\begin{proposition}\label{sec3 thm5}
		Suppose that $q$ is odd and $m$ is even. If $\gcd(h, q^m+1)\ge3$ or $\gcd(h+1, q^m+1)\ge 3$, then the minimum distance $d$ of $\mathcal{C}_{(q,q^m+1,3,h)}$ is equal to 4.
	\end{proposition}
	\begin{IEEEproof}
		By Corollary~\ref{sec3 cor1}, we have $d\ge4$ since $m$ is even. Suppose that $\gcd(h,q^m+1)=e\ge3$. Let $x=-1$, $y=1$, and $z\in U_{e}\setminus\{1,-1\}$. Let $w=-z\in U_{q^m+1}$. Since $\gcd(q-1,q^m+1)=2$, we have $z,w\notin\mathbb{F}_q$, i.e., $z^q-z\ne0$ and $w^q-w\ne0$. Thus, 
		\begin{equation*}
			\left(\frac{D(z,y)}{D(x,y)}\right)^q-\frac{D(z,y)}{D(x,y)}=\frac{1}{x^h}\left(\frac{(y-z)^q}{(y-x)^q}-\frac{y-z}{y-x}\right)=\frac{z-z^q}{2x^h}\ne0.
		\end{equation*}
		Similarly, we have 
		\begin{equation*}
			\left(\frac{D(x,z)}{D(x,y)}\right)^q-\frac{D(x,z)}{D(x,y)}=\frac{z^q-z}{2}\ne0.
		\end{equation*}
		Then one can easily check that
		\begin{equation*}
			\frac{\left(\frac{D(w,y)}{D(x,y)}\right)^q-\frac{D(w,y)}{D(x,y)}}{\left(\frac{D(z,y)}{D(x,y)}\right)^q-\frac{D(z,y)}{D(x,y)}}=\frac{\left(\frac{D(x,w)}{D(x,y)}\right)^q-\frac{D(x,w)}{D(x,y)}}{\left(\frac{D(x,z)}{D(x,y)}\right)^q-\frac{D(x,z)}{D(x,y)}}.
		\end{equation*}
		Take $l:=-1$ and
		\begin{equation*}
			k:=\frac{\left(\frac{D(w,y)}{D(x,y)}\right)^q-\frac{D(w,y)}{D(x,y)}}{\left(\frac{D(z,y)}{D(x,y)}\right)^q-\frac{D(z,y)}{D(x,y)}}=\frac{\left(\frac{D(x,w)}{D(x,y)}\right)^q-\frac{D(x,w)}{D(x,y)}}{\left(\frac{D(x,z)}{D(x,y)}\right)^q-\frac{D(x,z)}{D(x,y)}}.
		\end{equation*}
		Since $w=-z$, we have $k=-1$, and so $l,k\in\mathbb{F}^\ast_q$. 
		Let \begin{equation*}
			i:=k\frac{D(z,y)}{D(x,y)}+l\frac{D(w,y)}{D(x,y)}\ \mathrm{and}\ j:=k\frac{D(x,z)}{D(x,y)}+l\frac{D(x,w)}{D(x,y)}.
		\end{equation*}
		It is easily seen that $i^q-i=0$ and $j^q-j=0$, i.e., $i,j\in\mathbb{F}_q$. By Cramer's Rule, (\ref{sec3 thm2 equ2}) holds. Equivalently, (\ref{sec3 thm2 equ1}) is satisfied, and so $d=4$.
		
		The proof of the case where $\gcd(h + 1,q^m+1)\ge3$ is similar and thus omitted here. This completes the proof.
	\end{IEEEproof}
	
	\begin{remark}
		When $(q,m,h)=(5,2,2)$, we have $\gcd(h, q^m + 1)=2$ and $\gcd(h+1,q^m + 1)=1$. However, the minimum distance of $\mathcal{C}_{(q,q^m+1,3,h)}$ is 4 by Magma, which implies that the conditions in Proposition~\ref{sec3 thm5} are not necessary.
	\end{remark}
	
	\section{Two infinite families of distance-optimal linear codes}\label{sec4}
	
	In this section, we present two infinite families of distance-optimal linear codes and determine the parameters of $\mathcal{C}_{(q,q^m+1,3,h)}$ for certain $h$. Moreover, several linear codes with the best known parameters are given.
	
	\subsection{$q$ odd}
	
	Combining Theorem~\ref{sec3 thm4} and Proposition~\ref{sec3 thm5}, we obtain an infinite family of distance-optimal codes for $q$ odd.
	\begin{theorem}\label{sec3 thm6}
		Suppose that $q$ is odd and $h=$ 0 or $\frac{q^m+1}{2}$. Then the BCH code $\mathcal{C}_{(q,q^m+1,3,h)}$ has parameters $[q^m+1,q^m-2m,4]$. Specifically, when $q>3$, $\mathcal{C}_{(q,q^m+1,3,h)}$ is distance-optimal.
	\end{theorem}
	\begin{IEEEproof}
		When $h=$ 0 or $\frac{q^m+1}{2}$, $|C_{h}|=1$ and $|C_{h+1}|=2m$. Thus, the dimension of $\mathcal{C}_{(q,q^m+1,3,h)}$ is $q^m-2m$. Note that $\gcd(2h+1, q+1)=1$. If $m$ is odd, it follows from Theorem~\ref{sec3 thm4} that the minimum distance $d$ of $\mathcal{C}_{(q,q^m+1,3,h)}$ is 4. If $m$ is even, then $\gcd(h, q^m+1)\ge3$. According to Proposition~\ref{sec3 thm5}, $d=4$.
		
		If $d\ge5$, it then follows from the left side of (\ref{sec2 equ1}) that
		\begin{equation}\label{sec3 thm6 equ1}
			\sum\limits_{i=0}^{\lfloor\frac{d-1}{2}\rfloor}\binom{n}{i}(q-1)^i\ge1+(q^m+1)(q-1)+\frac{q^m(q^m+1)(q-1)^2}{2}.
		\end{equation}
		When $q>3$, we have $\sum\limits_{i=0}^{\lfloor\frac{d-1}{2}\rfloor}\binom{n}{i}(q-1)^i>q^{2m+1}$, which contradicts the Sphere Packing bound. That is, $d\le4$, and so $\mathcal{C}_{(q,q^m+1,3,h)}$ is distance-optimal. This completes the proof.
	\end{IEEEproof}
	
	\begin{remark}
		When $q=3$, and $m=1$ or $3$, the BCH code in Theorem~\ref{sec3 thm6} has the best known parameters according to the tables in \cite{Grassl}.
	\end{remark}
	
	The following theorem provides the minimum distance of the ternary BCH code $\mathcal{C}_{(3,3^m+1,3,h)}$ for some $m$ and $h$. 
	\begin{theorem}\label{sec3 thm7}
		Let $d$ be the minimum distance of the BCH code $\mathcal{C}_{(3,3^m+1,3,h)}$. Then
		\begin{equation*}
			d=\left\{\begin{array}{ll}
				3, \ &if\ m\ is\ odd,\ and\ \gcd(2h+1, 3^m+1)\ne1;\\
				4,\ &if\ m\ is\ odd\ and\ \gcd(2h+1, 3^m+1)=1,\ or\\
				&\quad m\ is\ even\ and\ either\ \gcd(h, 3^m+1)\ge3\ or\ \gcd(h+1, 3^m+1)\ge3;\\
				5,\ &if\ m\equiv2\bmod4,\ \gcd(h, 3^m+1)\le2\ and\ \gcd(h+1, 3^m+1)\le2;\\
				\ge5,\ &if\ m\equiv0\bmod4,\ \gcd(h, 3^m+1)\le2\ and\ \gcd(h+1, 3^m+1)\le2.
			\end{array} \right.
		\end{equation*}
	\end{theorem}
	\begin{IEEEproof}
		According to Theorem~\ref{sec3 thm4} and Proposition~\ref{sec3 thm5}, the cases of $d=$ 3 and 4 can be proved. We next show that $d\ge5$ if $m$ is even, $\gcd(h, 3^m+1)\le2$ and $\gcd(h+1, 3^m+1)\le2$. Without loss of generality, suppose that $h$ is odd and $h+1$ is even. Then  $\gcd(h,3^m+1)=1$ and $\gcd(h+1,3^m+1)=2$. Assume on the contrary that $d=4$. Then there exists $i,j,k,l\in\{1,-1\}\subset\mathbb{F}_3$ and four pairwise distinct elements $x,y,z,w\in U_{3^m+1}$ such that
		\begin{equation*}
			ix^{h}+jy^{h}+kz^{h}+lw^{h}=0,
		\end{equation*}
		and
		\begin{equation*}
			ix^{h+1}+jy^{h+1}+kz^{h+1}+lw^{h+1}=0.
		\end{equation*}
		The symmetry among $i,j,k$ and $l$ allows us to focus solely on the following cases:   
		\begin{itemize}
			\item $\boldsymbol{\{i,j,k,l\}=\{1,1,1,1\}}$.
			
			Then we get
			\begin{equation}\label{sec3 thm7 equ1}
				x^{h}+y^{h}+z^{h}+w^{h}=0,
			\end{equation}
			and
			\begin{equation}\label{sec3 thm7 equ2}
				x^{h+1}+y^{h+1}+z^{h+1}+w^{h+1}=0.
			\end{equation}
			Rising both sides of (\ref{sec3 thm7 equ1}) to $2$-th power, we have
			\begin{equation}\label{sec3 thm7 equ3}
				x^{2h}+y^{2h}+z^{2h}+w^{2h}=x^hy^h+x^hz^h+x^wy^h+y^hz^h+y^hw^h+z^hw^h.
			\end{equation}
			Rising both sides of (\ref{sec3 thm7 equ1}) to $3^m$-th power, we have
			\begin{equation}\label{sec3 thm7 equ4}
				x^{-h}+y^{-h}+z^{-h}+w^{-h}=0.
			\end{equation}
			Multiplying both sides of (\ref{sec3 thm7 equ4}) by $x^hy^hz^hw^h$ yields
			\begin{equation}\label{sec3 thm7 equ5}
				x^{h}y^{h}z^{h}+x^{h}y^{h}w^{h}+x^{h}z^{h}w^{h}+y^{h}z^{h}w^{h}=0.
			\end{equation}
			By multiplying (\ref{sec3 thm7 equ4}) and (\ref{sec3 thm7 equ5}), we obtain
			\begin{equation}\label{sec3 thm7 equ6}
				x^hy^h+x^hz^h+x^hy^h+y^hz^h+y^hw^h+z^hw^h=\frac{x^{2h}y^{2h}z^{2h}+x^{2h}y^{2h}w^{2h}+x^{2h}z^{2h}w^{2h}+y^{2h}z^{2h}w^{2h}}{x^hy^hz^hw^h}.
			\end{equation}
			Combining (\ref{sec3 thm7 equ3}) and (\ref{sec3 thm7 equ6}) leads to
			\begin{equation*}
				x^hy^hz^hw^h(x^{2h}+y^{2h}+z^{2h}+w^{2h})=x^{2h}y^{2h}z^{2h}+x^{2h}y^{2h}w^{2h}+x^{2h}z^{2h}w^{2h}+y^{2h}z^{2h}w^{2h}.
			\end{equation*}
			Rearranging terms,we have
			\begin{equation*}
				(x^hy^h-z^hw^h)(x^hz^h-y^hw^h)(x^hw^h-y^hz^h)=0.
			\end{equation*}
			Without loss of generality, assume that $x^hy^h=z^hw^h$. Multiplying both sides of (\ref{sec3 thm7 equ1}) by $x^h$ or $y^h$, we get
			\begin{equation*}
				\left\{\begin{array}{l}
					x^{2h}+x^hy^h+x^hz^h+x^hw^h=0,\\
					x^hy^h+y^{2h}+y^hz^h+y^hw^h=0.
				\end{array}\right.
			\end{equation*}
			Since $x^hy^h=z^hw^h$,
			\begin{equation}\label{sec3 thm7 equ7}
				x^hy^h+x^hz^h+x^wy^h+y^hz^h+y^hw^h+z^hw^h=-(x^{2h}+y^{2h}).
			\end{equation}
			Let $\sigma(u)=(u-x^h)(u-y^h)(u-z^h)(u-w^h)$. According to (\ref{sec3 thm7 equ1}), (\ref{sec3 thm7 equ5}) and (\ref{sec3 thm7 equ7}), we have
			\begin{equation*}
				\sigma(u)=u^4-(x^{2h}+y^{2h})u^2+x^{2h}y^{2h}=(u^2-x^{2h})(u^2-y^{2h}).
			\end{equation*}
			Then $x^h,y^h,-x^h$ and $-y^h$ are four roots of $\sigma(u)$ in $U_{3^m+1}$. Hence, either $x^h=-z^h$ and $y^h=-w^h$, or $x^h=-w^h$ and $y^h=-z^h$. Since $\gcd(h,3^m+1)=1$, we have either $x=-z$ and $y=-w$, or $x=-w$ and $y=-z$. Substituting $x=-z$ and $y=-w$ into (\ref{sec3 thm7 equ2}) leads to $z^{h+1}+w^{h+1}=0$, i.e., $(\frac{w}{z})^{2(h+1)}=1$. Note that $\gcd(2(h+1),3^m+1)=\gcd(4,3^m+1)=2$ as $m$ is even. That is, $\frac{w}{z}\in U_{2}=\{1,-1\}$. Since $z\ne w$, we have $z=-w=y$, a contradiction. On the other hand, substituting $x=-w$ and $y=-z$ into (\ref{sec3 thm7 equ2}) also leads to $z^{h+1}+w^{h+1}=0$, which is a contradiction.
			
			\item $\boldsymbol{\{i,j,k,l\}=\{1,1,1,-1\}}$.
			
			Let $w_1=-w$. Then we have
			\begin{equation}\label{sec3 thm7 equ8}
				x^{h}+y^{h}+z^{h}+w_1^{h}=0,
			\end{equation}
			and
			\begin{equation}\label{sec3 thm7 equ9}
				x^{h+1}+y^{h+1}+z^{h+1}-w_1^{h+1}=0.
			\end{equation}
			Note that $x,y,z$ and $w_1$ are different from each other. Otherwise, assume that $x=w_1$, it then follows form (\ref{sec3 thm7 equ9}) that $y^{h+1}+z^{h+1}=0$. Then $(\frac{y}{z})^{2h+2}=1$. Note that $\gcd(2h+2,3^m+1)=2$. Thus, we have $y=-z$, and so $2z^{h+1}=0$. This is impossible. Similar to the analysis of case where ${\{i,j,k,l\}=\{1,1,1,1\}}$, we have either $x=w$ or $y=w$, which is a contradiction.
			\item $\boldsymbol{\{i,j,k,l\}=\{1,1,-1,-1\}}$.
			
			Let $z_1=-z$ and $w_1=-w$.  Then we have
			\begin{equation}\label{sec3 thm7 equ10}
				x^{h}+y^{h}+z_1^{h}+w_1^{h}=0,
			\end{equation}
			and
			\begin{equation}\label{sec3 thm7 equ11}
				x^{h+1}+y^{h+1}-z_1^{h+1}-w_1^{h+1}=0.
			\end{equation}
			We claim that $x,y,z_1$ and $w_1$ are four pairwise distinct elements in $U_{3^m+1}$. If $x=w_1$, then $y^{h+1}-z_1^{h+1}=0$ by (\ref{sec3 thm7 equ11}), and so $y=z_1$. By (\ref{sec3 thm7 equ10}), we have $x^h+y^h=0$. Since $\gcd(h,3^m+1)=1$, $x=-y=z$, which contradicts $x\ne z$. It then follows from the proof of case where ${\{i,j,k,l\}=\{1,1,1,1\}}$ that either $x=z$ or $y=z$. This is impossible.
		\end{itemize}
		To sum up, we have $d\ge5$ if $m$ is even, $\gcd(h, 3^m+1)\le2$ and $\gcd(h+1, 3^m+1)\le2$.
		
		If $m\equiv 2 \bmod 4$, then $5 \mid 3^m+1$. Let $\alpha$ be the generator of $\mathbb{F}_{3^{2m}}$ and $\beta=\alpha^{3^m-1}$ be a primitive $(3^m+1)$-th root of unity in $\mathbb{F}_{3^{2m}}$. Note that $(\beta^h)^{3^m+1}-1=0$ and $(\beta^{h+1})^{3^m+1}-1=0$. Since $\gcd(h, 3^m+1)\le2$ and $\gcd(h+1, 3^m+1)\le2$, the order of the cyclic group $\left<\beta^h\right>$ or $\left<\beta^{h+1}\right>$ is at least $\frac{3^m+1}{2}$. Thus, $(\beta^h)^\frac{3^m+1}{5}-1\ne0$ and $(\beta^{h+1})^\frac{3^m+1}{5}-1\ne0$. Hence, we obtain
		\begin{equation*}
			\left\{\begin{aligned}
				&1+(\beta^h)^{\frac{3^m+1}{5}}+(\beta^h)^{\frac{2(3^m+1)}{5}}+(\beta^h)^{\frac{3(3^m+1)}{5}}+(\beta^h)^{\frac{4(3^m+1)}{5}}=0,\\
				&1+(\beta^{h+1})^{\frac{3^m+1}{5}}+(\beta^{h+1})^{\frac{2(3^m+1)}{5}}+(\beta^{h+1})^{\frac{3(3^m+1)}{5}}+(\beta^{h+1})^{\frac{4(3^m+1)}{5}}=0,
			\end{aligned}\right.
		\end{equation*}
		which indicates that $d=5$. This completes the proof.
	\end{IEEEproof}
	
	\begin{remark}
		1) Let $m=4$. The BCH codes $\mathcal{C}_{(3,3^m+1,3,3)}$ and $\mathcal{C}_{(3,3^m+1,3,4)}$ have minimum distance $6$ and $5$, respectively. Hence, the condition for $d=5$ in Theorem~\ref{sec3 thm7} is not necessary.
		
		2) If $h=0$, we have $\gcd(2h+1,3^m+1)=1$ and $\gcd(h,3^m+1)\ge3$. By Theorem~\ref{sec3 thm7}, the minimum distance of $\mathcal{C}_{(3,3^m+1,3,0)}$ is $4$, thereby proving the first part of Conjecture 23 in \cite{Li2017LCDCyclic}. Notably, our proof methods differ from those in \cite[Theorem IV.17]{Zhu2022AryAntiprimitive}.
	\end{remark}
	
	In the last case of Theorem~\ref{sec3 thm7}, we present a sufficient condition for the minimum distance of $\mathcal{C}_{(3,3^m+1,3,h)}$ to be greater than or equal to $6$.
	
	\begin{proposition}\label{sec3 pro4}
		Let $d$ be the minimum distance of the BCH code $\mathcal{C}_{(3,3^m+1,3,h)}$. Suppose that $m\equiv 0 \bmod 4$, $\gcd(h,3^m+1)\le2$, and $\gcd(h+1,3^m+1)\le2$. If there exists an integer $0\le i\le2m-1$ such that $(h+1)3^i \equiv 2h \bmod 3^m+1$, then $d\ge 6$.
	\end{proposition}
	\begin{IEEEproof}
		By Theorem~\ref{sec3 thm7}, we have $d\ge5$. Since $(h+1)3^i \equiv 2h \bmod 3^m+1$, $h+1$ is even. Thus, $\gcd(h,3^m+1)=1$ and $\gcd(h+1,3^m+1)=2$. For convenience, we introduce the notation $\sigma_{l,k}$ for the $k$-th elementary symmetric polynomial in $l$ variables $u_1^h,\dots,u_l^h$, defined by
		\begin{equation*}
			\sigma_{l,k}=\sigma_{l,k}(u_1^h,\dots,u_l^h)=\sum_{1\le i_1<\dots<i_k\le l}u_{i_1}^h\dots u_{i_k}^h.
		\end{equation*}
		Assume on the contrary that $d=5$. Then there exist $j_1,j_2,j_3,j_4,j_5\in\{1,-1\}\subset\mathbb{F}_3$ and five pairwise distinct elements $u_1,u_2,u_3,u_4,u_5\in U_{3^m+1}$ such that
		\begin{equation}\label{sec3 pro4 equ01}
			j_1u_1^{h}+j_2u_2^{h}+j_3u_3^{h}+j_4u_4^{h}+j_5u_5^{h}=0,
		\end{equation}
		and
		\begin{equation}\label{sec3 pro4 equ02}
			j_1u_1^{h+1}+j_2u_2^{h+1}+j_3u_3^{h+1}+j_4u_4^{h+1}+j_5u_5^{h+1}=0.
		\end{equation}
		Note that $(h+1)3^i \equiv 2h \bmod 3^m+1$. Thus, raising both sides of (\ref{sec3 pro4 equ02}) to $3^i$-th power leads to
		\begin{equation}\label{sec3 pro4 equ03}
			j_1u_1^{2h}+j_2u_2^{2h}+j_3u_3^{2h}+j_4u_4^{2h}+j_5u_5^{2h}=0.
		\end{equation}
		We only need to focus on the following cases due to the symmetry among $j_1,j_2,j_3,j_4,j_5$:
		\begin{itemize}
			\item $\boldsymbol{\{j_1,j_2,j_3,j_4,j_5\}=\{1,1,1,1,1\}}$.
			
			By (\ref{sec3 pro4 equ01}) and (\ref{sec3 pro4 equ03}), we have 
			\begin{equation}\label{sec3 pro4 equ1}
				u_1^{h}+u_2^{h}+u_3^{h}+u_4^{h}+u_5^{h}=\sigma_{5,1}=0,
			\end{equation}
			and
			\begin{equation}\label{sec3 pro4 equ2}
				u_1^{2h}+u_2^{2h}+u_3^{2h}+u_4^{2h}+u_5^{2h}=0.
			\end{equation}
			Raising both sides of (\ref{sec3 pro4 equ1}) to $2$-th power, we have $$u_1^{2h}+u_2^{2h}+u_3^{2h}+u_4^{2h}+u_5^{2h}+2\sigma_{5,2}=\sigma_{5,1}^2=0.$$
			By (\ref{sec3 pro4 equ2}), we have $\sigma_{5,2}=0$, and so $\sigma_{5,3}=\sigma_{5,5}\sigma_{5,2}^{3^m}=0$. Raising both sides of (\ref{sec3 pro4 equ1}) to $3^m$-th power, we have
			\begin{equation*}
				u_1^{-h}+u_2^{-h}+u_3^{-h}+u_4^{-h}+u_5^{-h}=0,
			\end{equation*}
			which implies that $\sigma_{5,4}=0$. Let $\sigma(x)=(x-u_1^h)(x-u_2^h)(x-u_3^h)(x-u_4^h)(x-u_5^h)$. Then $\sigma(x)=x^5-\sigma_{5,5}$. Note that $\gcd(5,3^m+1)=1$ since $m\equiv 0 \bmod 4$. Hence, $\sigma(x)$ can not have five different roots in $U_{3^m+1}$. 
			
			\item $\boldsymbol{\{j_1,j_2,j_3,j_4,j_5\}=\{1,1,1,1,-1\}}$.
			
			According to (\ref{sec3 pro4 equ01}) and (\ref{sec3 pro4 equ03}), we get
			\begin{equation}\label{sec3 pro4 equ4}
				u_1^{h}+u_2^{h}+u_3^{h}+u_4^{h}=\sigma_{4,1}=u_5^{h},
			\end{equation}
			and
			\begin{equation}\label{sec3 pro4 equ5}
				u_1^{2h}+u_2^{2h}+u_3^{2h}+u_4^{2h}=u_5^{2h}.
			\end{equation}
			Combining (\ref{sec3 pro4 equ4}) and (\ref{sec3 pro4 equ5}), we have $u_5^{2h}=\sigma_{4,1}^2-2\sigma_{4,2}$, and so $\sigma_{4,2}=0$. Raising both sides of (\ref{sec3 pro4 equ1}) to $3^m$-th power, we have
			\begin{equation}\label{sec3 pro4 equ6}
				u_1^{-h}+u_2^{-h}+u_3^{-h}+u_4^{-h}=u_5^{-h}.
			\end{equation}
			Multiplying both sides of (\ref{sec3 pro4 equ6}) by $\sigma_{4,4}$, we get
			$\sigma_{4,4}=u_5^h\sigma_{4,3}$. Let $\sigma(x)=(x-u_1^h)(x-u_2^h)(x-u_3^h)(x-u_4^h)$. Substituting $\sigma_{4,1}=u_5^h$, $\sigma_{4,2}=0$, and $\sigma_{4,4}=u_5^h\sigma_{4,3}$ into the expansion of $\sigma(x)$, we obtain $$\sigma(x)=x^4-u_5^hx^3-\sigma_{4,3}x+u_5^h\sigma_{4,3}=(x-u_5^h)(x^3-\sigma_{4,3}).$$ That is, $\sigma(x)$ has at most 2 roots in $U_{3^m+1}$, which is a contradiction.
			
			\item $\boldsymbol{\{j_1,j_2,j_3,j_4,j_5\}=\{1,1,1,-1,-1\}}$.
			
			Then (\ref{sec3 pro4 equ01}) and (\ref{sec3 pro4 equ03}) yield 
			\begin{equation}\label{sec3 pro4 equ7}
				u_1^{h}+u_2^{h}+u_3^{h}=\sigma_{3,1}=u_4^{h}+u_5^{h},
			\end{equation}
			and
			\begin{equation}\label{sec3 pro4 equ8}
				u_1^{2h}+u_2^{2h}+u_3^{2h}=u_4^{2h}+u_5^{2h}.
			\end{equation}
			Combining (\ref{sec3 pro4 equ7}) and (\ref{sec3 pro4 equ8}) leads to $\sigma_{3,2}=u_4^hu_5^h$, and so $\sigma_{3,2}\in U_{3^m+1}$. Note that $\sigma_{3,1}=\sigma_{3,3}\sigma_{3,2}^{q^m}$. Thus, we have $\sigma_{3,1}\in U_{3^m+1}$. By (\ref{sec3 pro4 equ7}), we obtain $u_4^{h}+u_5^{h}\in U_{3^m+1}$. Then
			\begin{equation}\label{sec3 pro4 equ9}
				1=(u_4^{-h}+u_5^{-h})(u_4^{h}+u_5^{h})=u_4^{-h}u_5^{h}+u_5^{-h}u_4^{h}-1.
			\end{equation}
			Multiplying both sides of (\ref{sec3 pro4 equ9}) by $u_4^hu_5^h$ yields
			\begin{equation*}
				u_4^{2h}+u_5^{2h}-2u_4^hu_5^h=0.
			\end{equation*}
			Then we have $(u_4^h-u_5^h)^2=0$, i.e., $u_4^h=u_5^h$. Since $u_4\ne u_5$ and $\gcd(h, 3^m+1)=1$, this is impossible.
		\end{itemize}
		Therefore, we have $d\ge6$. This proof is completed.
	\end{IEEEproof}
	
	\begin{remark}
		When $m\equiv4\bmod 8$, the minimum distance of $\mathcal{C}_{(3,3^m+1,3,1)}$ is 6 according to \cite[Theorem IV.25]{Zhu2022AryAntiprimitive}. Apart from this case, the results in \cite[Theorem~IV.25]{Zhu2022AryAntiprimitive} are contained in Theorem~\ref{sec3 thm7} and Proposition~\ref{sec3 pro4}.
	\end{remark}
	
	\subsection{$q$ even}
	
	When $q$ is even, the following theorem presents the parameters of $\mathcal{C}_{(q,q^m+1,3,h)}$ for specific values of $h$ and a family of distance-optimal linear codes.
	
	\begin{theorem}\label{sec3 thm8}
		Let $n=q^m+1$, where $q$ is even and $m$ is an integer. Let $1\le i\le 2m-1$ be an integer such that $\gcd(q^i-1,q^m+1)=1$. Let $h$ be the multiplicative inverse of $q^i-1$ modulo $n$. Then the BCH code $\mathcal{C}_{(q,n,3,h)}$ has parameters $[q^m+1,q^m+1-2m,d]$, where 
		\begin{equation*}
			d=\left\{\begin{array}{ll}
				3,\ &if\ m\ is\ odd;\\
				4,\ &if\ m\ is\ even\ and\ q>2;\\
				5,\ &if\ m \equiv 2 \bmod 4\ and\ q=2;\\
				5\ \mathrm{or}\ 6,\ &if\ m \equiv 0 \bmod 4\ and\ q=2.
			\end{array} \right.
		\end{equation*}
		Especially, if $m$ is even and $q>2$, $\mathcal{C}_{(q,n,3,h)}$ is distance-optimal.
	\end{theorem}
	\begin{IEEEproof}
		Since $h$ is the multiplicative inverse of $q^i-1$ modulo $n$, then we have $\gcd(h,n)=1$ and $h(q^i-1)\equiv 1 \bmod n$. Thus, we get  
		\begin{equation}\label{sec3 thm8 equ1}
			hq^i\equiv h+1 \bmod n,
		\end{equation}
		which implies that $h+1\in C_{h}$ and $C_{h}=C_{h+1}$. We first claim that $|C_h|=2m$. Assume that $|C_h|\ne 2m$. Then there exists $1\le j \le 2m-1$ such that $h(q^j-1)\equiv 0 \bmod n$, which contradicts $\gcd(h,n)=1$. Hence, the dimension of $\mathcal{C}_{(q,n,3,h)}$ equals $q^m+1-2m$. The minimum distance of $\mathcal{C}_{(q,n,3,h)}$ is analyzed as follows.
		
		If $m$ is odd, then $q+1 \mid q^m+1$, and so $2h+1\equiv h(q^i+1) \bmod {q+1}$ by (\ref{sec3 thm8 equ1}). Then $\gcd(2h+1,q+1)=\gcd(h(q^i+1),q+1)=\gcd(q^i+1,q+1)$ as $\gcd(h, q+1)=1$. Note that $\gcd(q^i-1,q^m+1)=1$ if and only if $q$ is even and $\frac{i}{\gcd(i,m)}$ is odd. Since $m$ is odd, we have $\gcd(i,m)$ is odd, and so $i$ is odd. Thus, $\gcd(2h+1,q+1)=q+1\ne1$. It then follows from Corollary~\ref{sec3 cor1} that $d=3$.
		
		If $m$ is even, then $d\ge4$ by Corollary~\ref{sec3 cor1}. 
		\begin{itemize}
			\item $\boldsymbol{q>4}$. 
			
			Assume on the contrary that $d\ge5$, it then follows from the left side of (\ref{sec2 equ1}) that
			\begin{equation*}
				\sum\limits_{i=0}^{\lfloor\frac{d-1}{2}\rfloor}\binom{n}{i}(q-1)^i\ge1+(q^m+1)(q-1)+\frac{q^m(q^m+1)(q-1)^2}{2}>q^{2m},
			\end{equation*}
			By the Sphere Packing bound, this is impossible. Hence, we have $d=4$.
			\item $\boldsymbol{q=2}$. 
			
			Assume on the contrary that $d\ge7$. It follows from the left side of (\ref{sec2 equ1}) that
			\begin{equation*}
				\sum\limits_{i=0}^{\lfloor\frac{d-1}{2}\rfloor}\binom{n}{i}(q-1)^2\ge1+(2^m+1)+2^{m-1}(2^m+1)+\frac{2^{m-1}(2^{2m}-1)}{3}>2^{2m},
			\end{equation*}
			which contradicts the Sphere Packing bound. Hence, we have $4\le d\le6$. We claim that $d\ge5$. Assume that $d=4$, then there is a codeword in $\mathcal{C}_{(2,n,3,h)}$ with weight 4. That is, there exist three pairwise distinct elements $x,y,z\in U_{2^m+1}\setminus\{1\}$ such that
			\begin{equation}\label{sec3 thm8 equ2}
				1+x^h+y^h+z^h=0.
			\end{equation}
			Raising both side of (\ref{sec3 thm8 equ2}) to the $2^m$-th power, we have
			\begin{equation}\label{sec3 thm8 equ3}
				1+x^{-h}+y^{-h}+z^{-h}=0.
			\end{equation}
			Combining (\ref{sec3 thm8 equ2}) and (\ref{sec3 thm8 equ3}) leads to
			\begin{equation*}
				1=(1+x^h+y^h)(1+x^{-h}+y^{-h})=1+x^h+y^h+x^{-h}+y^{-h}+x^hy^{-h}+x^{-h}y^{h},
			\end{equation*}
			which yields
			\begin{equation}\label{sec3 thm8 equ4}
				x^h+y^h+x^{-h}+y^{-h}=x^hy^{-h}+x^{-h}y^{h}.
			\end{equation}
			By (\ref{sec3 thm8 equ2}) and (\ref{sec3 thm8 equ3}), we have
			\begin{equation}\label{sec3 thm8 equ5}
				x^h+y^h+x^{-h}+y^{-h}=z^h+z^{-h}.
			\end{equation}
			Hence, it follows from (\ref{sec3 thm8 equ4}) and (\ref{sec3 thm8 equ5}) that
			\begin{equation*}
				x^hy^{-h}+x^{-h}y^{h}=z^h+z^{-h},
			\end{equation*}
			which implies that 
			\begin{equation*}
				(z^h+x^hy^{-h})(z^{h}+x^{-h}y^h)=0.
			\end{equation*}
			That is, $z^h=x^hy^{-h}$ or $z^h=x^{-h}y^h$. Since $\gcd(h, q^m+1)=1$ and $q$ is even, $a^h+b^h\ne0$ for any two different elements $a,b\in U_{q^m+1}$. Substituting $z^h=x^hy^{-h}$ into (\ref{sec3 thm8 equ2}), we have $(x^h+y^h)(1+y^{-h})=0$, which is a contradiction. The case of $z^h=x^{-h}y^h$ is similar. Therefore,  we have $5\le d\le6$ if $m$ is even.

			If $m \equiv 2 \bmod 4$, then $5 \mid 2^m+1$. Let $\alpha$ be the generator of $\mathbb{F}_{2^{2m}}$ and $\beta=\alpha^{2^m-1}$ be a primitive $(2^m+1)$-th root of unity in $\mathbb{F}_{2^{2m}}$. Since $\gcd(h, 2^m+1)=1$, we have $(\beta^h)^{2^m+1}=1$ and $(\beta^h)^{\frac{2^m+1}{5}}\ne1$. Hence,  
			\begin{equation*}
				1+(\beta^h)^{\frac{2^m+1}{5}}+(\beta^h)^{\frac{2(2^m+1)}{5}}+(\beta^h)^{\frac{3(2^m+1)}{5}}+(\beta^h)^{\frac{4(2^m+1)}{5}}=0,
			\end{equation*}
			which indicates that $d=5$. This completes the proof.
		\end{itemize}
	\end{IEEEproof}
	
	\begin{example}
		1) When $m=4, 8$ or $12$, the BCH code $\mathcal{C}_{(2,2^m+1,3,h)}$ in Theorem~\ref{sec3 thm8} has parameters $[2^m+1, 2^m+1-2m, 5]$ through Magma experiments. Hence, we conjecture that the minimum distance of $\mathcal{C}_{(2,2^m+1,3,h)}$ in Theorem~\ref{sec3 thm8} is 5 when $m\equiv0\bmod 4$.
		
		2) When $m=2,4$ or $6$, the BCH code $\mathcal{C}_{(2,2^m+1,3,h)}$ in Theorem~\ref{sec3 thm8} has the best known parameters according to \cite{Grassl}.
	\end{example}
	
	\begin{remark}
		1) When $m=2$ and $q>2$, the BCH code $\mathcal{C}_{(q,q^m+1,3,h)}$ in Theorem~\ref{sec3 thm8} is an AMDS code with parameters $[q^2+1,q^2-3,4]$.
		
		2) Notably, the BCH code $\mathcal{C}_{(q,q^m+1,3,h)}$ in Theorem~\ref{sec3 thm8} is equivalent to $\mathcal{C}_{(q,q^m+1,2,h)}$. The parameters of $\mathcal{C}_{(q,q^m+1,2,1)}$ were partially given in \cite[Theorem IV.22]{Zhu2022AryAntiprimitive}, and they coincide with those provided in Theorem~\ref{sec3 thm8}. When $q\ge4$, the value $h=1$ violates the conditions of Theorem~\ref{sec3 thm8}; however, for $q=2$, the case of $h=1$ falls under the scope of Theorem~\ref{sec3 thm8}.
	\end{remark}
	
	The following theorem presents the further exploration of the case where $h=1$ and $q=2$.
	
	\begin{theorem}\label{sec3 thm9}
		Let $m\equiv0\bmod4$ be an integer. The the BCH code $\mathcal{C}_{(2,2^m+1,3,1)}$ has parameters $[2^m+1,2^m+1-2m,d]$, where 
		\begin{equation*}
			d=\left\{\begin{array}{ll}
				5,\ &if\ m \not\equiv 0 \bmod 16;\\
				5\ \mathrm{or}\ 6,\ &if\ m \equiv 0 \bmod 16.
			\end{array} \right.
		\end{equation*}
	\end{theorem}
	\begin{IEEEproof}
		Let $\alpha$ be the generator of $\mathbb{F}_{2^{2m}}$ and $\beta=\alpha^{2^m-1}$ be a primitive $(2^m+1)$-th root of unity in $\mathbb{F}_{2^{2m}}$. According to Theorem~\ref{sec3 thm8}, we only need to prove that $d=5$ when $m \not\equiv 0 \bmod 16$. Since $m\equiv0\bmod4$, we need check the cases of $m\equiv4\bmod8$ and $m\equiv8\bmod16$. 
		
		If $m\equiv4\bmod8$, then $2^4 + 1\mid2^m + 1$, and so $x^{17}-1\mid x^{2^m+1}-1$. Let $\gamma=\beta^{\frac{2^m+1}{17}}$ be a primitive $17$-th root of unity in $\mathbb{F}_{2^{2m}}$. Denote the $2$-cyclotomic coset of $1$ and $3$ modulo $17$ by $C_{1,17}$ and $C_{3,17}$, respectively. Let $f_1(x)=\sum_{i\in\{0\}\cup C_{1,17}}x^i$ and $f_3(x)=\sum_{i\in\{0\}\cup C_{3,17}}x^i$. It is easily checked that $\mathbb{Z}_{17}=\{0\}\cup C_{1,17}\cup C_{3,17}$. Then
		\begin{equation}\label{sec3 thm9 equ1}
			0=\sum_{i\in\mathbb{Z}_{17}}\gamma^i=1+f_1(\gamma)+f_3(\gamma).
		\end{equation}
		Since $f_1(\gamma)^2=f_1(\gamma)$ and $f_3(\gamma)^2=f_3(\gamma)$, we have $f_1(\gamma),f_3(\gamma)\in \{0,1\}$. By (\ref{sec3 thm9 equ1}), either $f_1(\gamma)=0$ or $f_3(\gamma)=0$. Note that $f_1(x)=f_3(x^3)$. Hence, either $f_3(\gamma^3)=0$ or $f_3(\gamma)=0$. Let $f(x)=\gcd\left(x^{17}-1,f_3(x)\right)=x^8 + x^5 + x^4 + x^3 + 1.$ Then we get that either $f(\gamma^3)=0$ or $f(\gamma)=0$. Since $f(x)$ is a five-term polynomial over $\mathbb{F}_2$, $\mathcal{C}_{(2,2^m+1,3,1)}$ has a codeword with weight $5$.
		
		If $m\equiv8\bmod16$, then $2^8 + 1\mid2^m + 1$, and so $x^{257}-1\mid x^{2^m+1}-1$. Let $\gamma=\beta^{\frac{2^m+1}{257}}$ be a primitive $257$-th root of unity in $\mathbb{F}_{2^{2m}}$. Note that there exists a six-term polynomial $f(x)=x^{16} + x^{15} + x^8 + x + 1 \mid x^{257}-1$. Moreover, one can verify that $\gamma^7$ is a root of $f(x)$ by Magma. Then we have $d=5$. The proof is completed.
	\end{IEEEproof}
	
	\begin{remark}
		According to Theorems~\ref{sec3 thm8} and \ref{sec3 thm9}, more precise minimum distances of $\mathcal{C}_{(q,q^m+1,2,1)}$ are derived, thereby improving the results of Theorem IV.22 in \cite{Zhu2022AryAntiprimitive}.
	\end{remark}
	
	In \cite{Xu2024Thedual}, an infinite family of MDS codes was derived from the BCH code $\mathcal{C}_{(q^m,q^m+1,3,h)}$ for $q$ even.
	\begin{corollary}\cite[Corollary~1]{Xu2024Thedual}\label{sec3 cor2}
		Let $q=2^s$ and $h=\frac{q^m-2^i}{2}$, where $s>0$, $m>1$ and $0<i<sm$ are three integers. If $\gcd(i,sm)=1$, then the BCH code $\mathcal{C}_{(q^m,q^m+1,3,h)}$ is an MDS code with parameters $[q^m+1,q^m-3,5]$.
	\end{corollary}
	
	Note that the subfield subcode of $\mathcal{C}_{(q^m,q^m+1,3,h)}$ with respect to $\mathbb{F}_q$ is $\mathcal{C}_{(q,q^m+1,3,h)}$. We next consider the parameters of subfield subcode of $\mathcal{C}_{(q^m,q^m+1,3,h)}$ in Corollary~\ref{sec3 cor2}.
	\begin{theorem}\label{sec3 thm10}
		Let $q=2^s$ and $h=\frac{q^m-2^i}{2}$, where $s>0$, $m>1$ and $0<i<sm$ are three integers. If $\gcd(i,sm)=1$, then the BCH code $\mathcal{C}_{(q,q^m+1,3,h)}$ has parameters 
		\begin{equation*}
			[n,k,d]=\left\{\begin{array}{ll}
				[q^m+1,1,q^m+1],\ &if\ (s, m, i)=(1,2,1),\ (1,3,1)\ or\ (1,3,2);\\ \relax
				[q^m+1,q^m + 1 - 4m,5],\ &if\ sm \equiv 2 \bmod 4;\\ \relax
				[q^m+1,q^m + 1 - 4m,d\ge5]\ &otherwise.\\
			\end{array} \right.
		\end{equation*}
	\end{theorem}
	\begin{IEEEproof}
		The dimension of $\mathcal{C}_{(q,q^m+1,3,h)}$ is $q^m+1-|C_h\cup C_{h+1}|$. To determine the value of $k$, we will prove the following three proposition:
		\begin{enumerate}
			\item $|C_h|<2m$ if and only if $(s,m, i)=(1,3,1)$;
			\item $|C_{h+1}|<2m$ if and only if $(s,m, i)=(1,3,2)$;
			\item $C_h=C_{h+1}$ if and only if $(s,m, i)=(1,2,1)$.
		\end{enumerate}
		Combining 1), 2) and 3), we have $k=q^m+1-4m$ if and only if $(s,m,i)\notin\{(1,3,1),(1,3,2),(1,2,1)\}$. Hence, the dimension of $\mathcal{C}_{(q,q^m+1,3,h)}$ is determined. When $(s,m,i)=(1,3,1),(1,3,2)$ or $(1,2,1)$, one can easily check that $\mathcal{C}_{(q,q^m+1,3,h)}$ is a trivial linear code with parameters $[q^m+1,1,q^m+1]$.
		
		1) We now prove that $|C_h|<2m$ if and only if $(s,m, i)=(1,3,1)$. Suppose that $|C_h|=\ell$. Then  $\ell \mid 2m$ by Theorem~\ref{sec2 thm2}, and $h\equiv hq^\ell \bmod q^m+1$, i.e.,
		\begin{equation}\label{sec4 thm10 equ0}
			\frac{2^{sm}-2^i}{2}2^{s\ell}\equiv\frac{2^{sm}-2^i}{2}\bmod 2^{sm}+1.
		\end{equation}
		Multiplying both sides of (\ref{sec4 thm10 equ0}) by 2, we get
		\begin{equation*}
			2^{sm}-2^i\equiv(2^{sm}-2^i)2^{s\ell}\bmod 2^{sm}+1.
		\end{equation*}
		which is the same as 
		\begin{equation*}
			(2^i+1)(2^{s\ell}-1)\equiv0\bmod 2^{sm}+1.
		\end{equation*}
		That is, 
		\begin{equation}\label{sec4 thm10 equ1}
			2^{sm}+1 \mid (2^i+1)(2^{s\ell}-1).
		\end{equation}
		If $(s,m, i)=(1,3,1)$, it is easy to check that $\ell<2m$. Conversely, if $\ell<2m$, then $\ell\le \frac{1}{2}\times2m=m$, and so $2^{sm}+1\nmid2^{s\ell}-1$ and $\gcd(2^i+1,2^sm+1)>1$. According to Lemma~\ref{sec2.3 lem1} and $\gcd(i,sm)=1$, we have
		\begin{equation}\label{sec4 thm10 equ2}
			\gcd(2^{i}+1,2^{sm}+1)=3,
		\end{equation}
		and $i$, $m$ and $s$ are odd. If $\ell\mid m$, then $\gcd(2^{s\ell}-1,2^{sm}+1)=1$ by Lemma~\ref{sec2.3 lem2}, which indicates that $2^{sm}+1\mid2^i+1$ by (\ref{sec4 thm10 equ1}). This is impossible as $i<sm$. Thus, we have $\ell\nmid m$. Since $\ell\mid2m$ and $\ell\nmid m$, we get $\gcd(\ell,m)=\frac{\ell}{2}$. By Lemma~\ref{sec2.3 lem2}, we have
		\begin{equation}\label{sec4 thm10 equ3}
			\gcd(2^{s\ell}-1,2^{sm}+1)=2^{s\frac{\ell}{2}}+1.
		\end{equation}  
		Combining (\ref{sec4 thm10 equ1}), (\ref{sec4 thm10 equ2}) and (\ref{sec4 thm10 equ3}) yields $2^{sm}+1 \mid 3(2^{s\frac{\ell}{2}}+1)$. Suppose that $m=\frac{\ell}{2}j$, then 
		\begin{equation*}
			\begin{aligned}
				2^{sm}+1&=(2^{s\frac{\ell}{2}}+1)(2^{{s\frac{\ell}{2}}(j-1)}-2^{{s\frac{\ell}{2}}(j-2)}+\dots-2^{s\frac{\ell}{2}}+1)\\
				&=(2^{s\frac{\ell}{2}}+1)\big((2^{s\frac{\ell}{2}}-1)(2^{{s\frac{\ell}{2}}(j-2)}+2^{{s\frac{\ell}{2}}(j-4)}+\dots+2^{{s\frac{\ell}{2}}})+1\big).
			\end{aligned}
		\end{equation*}
		Thus, we have $\big(2^{s\frac{\ell}{2}}-1)(2^{{s\frac{\ell}{2}}(j-2)}+2^{{s\frac{\ell}{2}}(j-4)}+\dots+2^{{s\frac{\ell}{2}}})+1\big) \mid 3$, which leads to $s=1$, $l=2$ and $m=j=3$. Since $\gcd(i, sm)=1$ and $i$ is odd, $i=1$. That is, $(s,m, i)=(1,3,1)$.
		
		2) Now suppose that $|C_{h+1}|=\ell$, then $2^{sm}+1 \mid (2^i-1)(2^{s\ell}-1)$. Through a very similar analysis, one can obtain that $|C_{h+1}|<2m$ if and only if $(s,m,i)=(1,3,2)$.
		
		3) We next show that $C_h=C_{h+1}$ if and only if $(s,m,i)=(1,2,1)$. If $(s,m,i)=(1,2,1)$, then $h=1\in C_{h+1}$, i.e., $C_h=C_{h+1}$. If $C_h=C_{h+1}$, then there exist an integer $1\le j\le2m-1$ such that
		\begin{equation*}
			\frac{2^{sm}-2^i}{2}2^{sj}\equiv\frac{2^{sm}-2^i}{2}+1\bmod 2^{sm}+1,
		\end{equation*}
		which indicates that $2^{sm}+1 \mid (2^i+1)(2^{sj}-1)+2$. Note that $i<sm$. Assume on the contrary that $j\ge m$, then
		\begin{equation*}
			(2^i+1)(2^{sj}-1)+2=2^{sj-sm}(2^i+1)(2^{sm}+1)-2^{sj+i-sm}-2^{sj-sm}-2^i+1.
		\end{equation*}
		If $sj+i-sm< sm$, then we have $-2^{sj+i-sm}-2^{sj-sm}-2^i+1=0$, which is impossible as $i>0$. If $sj+i-sm\ge sm$, then
		\begin{equation*}
			-2^{sj+i-sm}-2^{sj-sm}-2^i+1=-2^{sj+i-2sm}(2^{sm}+1)+2^{sj+i-2sm}-2^{sj-sm}-2^i+1.
		\end{equation*}
		Since $j<2m$ and $i<sm$, we have $sj+i-2sm<sm$ and $sj-sm<sm$, and so $2^{sj+i-2sm}-2^{sj-sm}-2^i+1=0$. Thus, $2^{sj+i-2sm}+1=2^{sj-sm}+2^i$, which leads to either $sj+i-2sm=i$ or $sj+i-2sm=sj-sm$. That is, either $j=2m$ or $i=sm$, a contradiction. Hence, we have $j<m$. Then 
		\begin{equation*}
			(2^i+1)(2^{sj}-1)+2=2^{sj+i-sm}(2^{sm}+1)-2^{sj+i-sm}+2^{sj}-2^i+1.
		\end{equation*}
		Since $j<m$ and $i<sm$, we have $sj+i-sm<sm$. Then $-2^{sj+i-sm}+2^{sj}-2^i+1=0$, which leads to $sj=i$ and $sj+i-sm=0$. Then we get $i=sj=\frac{sm}{2}$. Since $\gcd(i,sm)=1$, $i=1,s=1$ and $m=2$. That is, $(s,m,i)=(1,2,1)$. 
		
		By Corollary~\ref{sec3 cor2} and the definition of subfield subcodes, we have $d\ge5$. Note that
		\begin{equation*}
			\frac{2^{sm}-2^i}{2}\equiv2^{sm-1}(2^i-2^{sm})\equiv2^{sm-1}(2^i+1)\bmod2^{sm}+1,
		\end{equation*}
		and
		\begin{equation*}
			\frac{2^{sm}-2^i}{2}+1\equiv2^{sm-1}(2^i+1)-2^{sm}\equiv2^{sm-1}(2^i-1)\bmod 2^{sm}+1.
		\end{equation*}
		Then we have $\gcd(h, q^m+1)==\gcd(2^i+1,2^{sm}+1)$ and $\gcd(h+1, q^m+1)=\gcd(2^i-1,2^{sm}+1)$. Since  $\gcd(i,sm)=1$, we have $\gcd(h, 2^{sm}+1),\gcd(h+1, 2^{sm}+1)\in\{1,3\}$ by Lemmas~\ref{sec2.3 lem1} and \ref{sec2.3 lem2}. Let $\beta$ be a primitive $q^m+1$-th root of unity in $\mathbb{F}_{q^{2m}}$. Then the orders pf the cyclic groups $\left<\beta^h\right>$ and $\left<\beta^{h+1}\right>$ are at least $\frac{q^m+1}{3}$. Hence, $(\beta^h)^\frac{q^m+1}{5}-1\ne0$ and $(\beta^{h+1})^\frac{q^m+1}{5}-1\ne0$.  If $sm \equiv 2 \bmod 4$, then $5 \mid 2^{sm}+1$. Since $(\beta^{h})^{q^m+1}-1=0$ and $(\beta^{h+1})^{q^m+1}-1=0$,
		\begin{equation*}
			\left\{\begin{aligned}
				&1+(\beta^h)^{\frac{q^m+1}{5}}+(\beta^h)^{\frac{2(q^m+1)}{5}}+(\beta^h)^{\frac{3(q^m+1)}{5}}+(\beta^h)^{\frac{4(q^m+1)}{5}}=0,\\
				&1+(\beta^{h+1})^{\frac{q^m+1}{5}}+(\beta^{h+1})^{\frac{2(q^m+1)}{5}}+(\beta^{h+1})^{\frac{3(q^m+1)}{5}}+(\beta^{h+1})^{\frac{4(q^m+1)}{5}}=0.
			\end{aligned}\right.
		\end{equation*}
		Therefore, $d=5$ if $sm \equiv 2 \bmod 4$. This proof is completed.
	\end{IEEEproof}
	
	\begin{example}
		Here we present the following examples for the BCH code of Theorem~\ref{sec3 thm10}. Suppose that $h=\frac{q^m-2}{2}$.
		\begin{itemize}
			\item The code $\mathcal{C}_{(2,2^5+1,3,h)}$ has parameters $[33,13,10]$.
			\item The code $\mathcal{C}_{(4,4^2+1,3,h)}$ has parameters $[17,9,7]$.
			\item The code $\mathcal{C}_{(8,8^2+1,3,h)}$ has parameters $[65,57,5]$.
		\end{itemize}
		The above three codes have the best known parameters by the table of \cite{Grassl}.
	\end{example}
	
	\section{Concluding remarks}\label{sec5}
	
	For any $q$ and $m$, we established the sufficient and necessary condition for the minimum distance $d$ of the antiprimitive BCH code $\mathcal{C}_{(q,q^m+1,3,h)}$ to be $3$. For odd $q$ and $m$, we provided the sufficient and necessary condition for $d=4$ and completely determined the value of $d$. For the cases where $q$ or $m$ is even, we presented some sufficient or necessary conditions for $d=4$. According to these results, we derived the parameters of $\mathcal{C}_{(q,q^m+1,3,h)}$ for certain $h$. Additionally, two infinite families of distance-optimal codes and several linear codes with the best known parameters were obtained. Notably, our results extend the related findings in \cite[Theorems~1 and 3]{Xu2025TheSufficient} and \cite[Theorem IV.22]{Zhu2022AryAntiprimitive}.
	
	When $q$ or $m$ is even, it is not easy to derive a concise sufficient and necessary condition for the minimum distance of code $\mathcal{C}$ to be $4$ through the proof techniques of Theorems~\ref{sec3 thm1} and \ref{sec3 thm4}. Thus, the existence of alternative methods for determining the minimum distance requires further investigation.

	\renewcommand\refname{References}

\end{document}